\documentclass[conference]{IEEEtran}
\IEEEoverridecommandlockouts
\usepackage{cite}
\usepackage{amsmath,amssymb,amsfonts}
\usepackage{algorithmic}
\usepackage{graphicx}
\usepackage{textcomp}
\usepackage{xcolor}
\usepackage{float}
\usepackage{threeparttable} 
\usepackage{soul}
\usepackage{hyperref}

\def\BibTeX{{\rm B\kern-.05em{\sc i\kern-.025em b}\kern-.08em
    T\kern-.1667em\lower.7ex\hbox{E}\kern-.125emX}}

\begin{document} 
\title{An Optical XNOR-Bitcount Based Accelerator for Efficient Inference of Binary Neural Networks}

\author{\IEEEauthorblockN{Sairam Sri Vatsavai}
\IEEEauthorblockA{\textit{Electrical and Computer Engineering} \\
\textit{University of Kentucky}\\
Lexington, USA \\
ssr226@uky.edu}
\and
\IEEEauthorblockN{Venkata Sai Praneeth Karempudi }
\IEEEauthorblockA{\textit{Electrical and Computer Engineering} \\
\textit{University of Kentucky}\\
Lexington, USA  \\
kvspraneeth@uky.edu}
\and
\IEEEauthorblockN{Ishan Thakkar }
\IEEEauthorblockA{\textit{Electrical and Computer Engineering} \\
\textit{University of Kentucky}\\
Lexington, USA  \\
igthakkar@uky.edu}
}

\maketitle

\begin{abstract}
Binary Neural Networks (BNNs) are increasingly preferred over full-precision Convolutional Neural Networks (CNNs) to reduce the memory and computational requirements of inference processing with minimal accuracy drop. BNNs convert CNN model parameters to 1-bit precision, allowing inference of BNNs to be processed with simple XNOR and bitcount operations. This makes BNNs amenable to hardware acceleration. Several photonic integrated circuits (PICs) based BNN accelerators have been proposed. Although these accelerators provide remarkably higher throughput and energy efficiency than their electronic counterparts, the utilized XNOR and bitcount circuits in these accelerators need to be further enhanced to improve their area, energy efficiency, and throughput. This paper aims to fulfill this need. For that, we invent a single-MRR-based optical XNOR gate (OXG). Moreover, we present a novel design of bitcount circuit which we refer to as Photo-Charge Accumulator (PCA). We employ multiple OXGs in a cascaded manner using dense wavelength division multiplexing (DWDM) and connect them to the PCA, to forge a novel \underline{\textbf{O}}ptical \underline{\textbf{X}}NOR-Bitcount based \underline{\textbf{B}}inary \underline{\textbf{N}}eural \underline{\textbf{N}}etwork \underline{\textbf{A}}ccelerator (OXBNN). 
Our evaluation for the inference of four modern BNNs indicates that OXBNN provides improvements of up to 62$\times$ and  7.6$\times$ in frames-per-second (FPS) and FPS/W (energy efficiency), respectively, on geometric mean over two PIC-based BNN accelerators from prior work. We developed a transaction-level, event-driven python-based simulator for evaluation of accelerators (\url{https://github.com/uky-UCAT/B\_ONN\_SIM})\footnote{To Appear at IEEE ISQED 2023}. 
\end{abstract}

\section{Introduction}
Convolutional Neural Networks (CNNs) have revolutionized the implementation of various artificial intelligence tasks, such as image recognition, language translation, and autonomous driving \cite{dnnapplications1,dnnapplications2}, due to their high inference accuracy. However, the heavy computation and storage requirements of CNNs still limit their application in practice. Therefore, to improve the speed and efficiency of CNN inference, model compression techniques such as quantization are widely employed \cite{gupta2015deep,zhu2020towards,gong2014compressing}. Quantization techniques create compact CNNs compared to their floating-point counterparts by representing the weights/inputs of CNNs with lower precision. The extreme end of the quantization is binarization, i.e., a 1-bit quantization, that allows only two possible values for both inputs and weights, either -1(0) or +1.

Binarization replaces the heavy floating-point vector-dot-product operations (which constitute convolution operations in CNNs) with simple bit-wise XNOR and bitcount operations \cite{rastegari2016xnor}. Since bit-wise XNOR and bitcount are lightweight operations, binarized CNNs, referred to as binary neural networks (BNNs), provide efficient hardware implementations. Among the BNN hardware implementations from prior works, the silicon-photonic accelerators have shown great promise to provide unparalleled parallelism, ultra-low latency, and high energy efficiency \cite{lightbulb,robin}. Prior work \cite{lightbulb} utilizes microdisks to realize XNOR-Bitcount processing cores (XPCs) that process the input and weight vectors, whereas \cite{robin} uses Microring Resonators (MRRs) in its XPCs to perform XNOR-Bitcount operations. However, these prior works face two shortcomings. First, they use at least two MRRs or microdisks to achieve 1-bit XNOR operation, which increases their area and energy consumption. Second, because of the limited scalability of their XNOR and bitcount circuits, they are forced to decompose the input and weight vectors into a large number of smaller slices before processing them. This generates a large number of partial sums (\textit{psum}s). Accumulating such a large number of \textit{psum}s to obtain the final result, using a \textit{psum} reduction network, can incur a very high latency overhead.

To address these shortcomings, this paper presents a novel Optical XNOR-Bitcount based Binary Neural Network Accelerator (OXBNN). OXBNN employs a novel design of optical XNOR gates (OXGs). Our OXG uses a single MRR to perform a 1-bit XNOR operation, thereby reducing the area and energy consumption compared to prior works. Moreover, OXBNN employs a novel bitcount circuit, referred to as Photo-Charge Accumulator (PCA), which inherently supports the accumulation of a very high number of \textit{psum}s, thereby eliminating the need of using external \textit{psum} reduction networks, to consequently reduce the overall latency and energy consumption of BNN processing.          

Our key contributions in this paper are summarized below.
\begin{itemize}
    \item We present our invented, novel BNN accelerator called OXBNN, which employs an array of single-MRR-based optical XNOR gates (OXGs) and highly scalable bitcount circuits called Photo-Charge Accumulators (PCAs);
    \item  We perform detailed modeling and characterization of our invented OXGs and PCAs using photonics foundry-validated, commercial-grade, photonic-electronic design automation tools (Section III);

\item We perform a scalability analysis for
our OXBNN and describe a pertinent mapping scheme (Section IV);

\item We implement and evaluate OXBNN at the system-
level with our in-house simulator (\url{https://github.com/uky-UCAT/B\_ONN\_SIM}), and compare its performance with two well-known photonic BNN accelerators from prior works, for the inferences of
four state-of-the-art BNNs (Section V).

\end{itemize}
\section{Preliminaries}
\subsection{Binary Neural Networks (BNNs)}\label{section2a}
BNNs are specific types of CNNs that employ quantization techniques \cite{lqnets} to quantize the weights and inputs to 1-bit values, reducing the storage requirements and computational effort for improved energy efficiency of model inference. With binary quantization, the weights and inputs can only assume two possible values, either -1 or 1  \cite{courbariaux2016binarized,rastegari2016xnor}. In general, the \textit{sign} function is the most widely used binary quantization function (Q):
\begin{equation}
    Q(x) = sign(x) = x \geq 0 \; ? +1 : -1 
\end{equation}

Like for CNNs \cite{deapcnn}, a convolution operation for BNNs is also typically decomposed into multiple vector-dot-product (VDP) operations. Each VDP operation of a BNN occurs between two vectors, the individual elements of which are first binarized using Eq. (1). Then, the VDP operation between a binarized weight vector \textit{W} and a binarized input vector \textit{I} can be realized in two steps, in this given order: (i) element-wise (i.e., bit-wise) XNOR of \textit{I} and \textit{W} that produces an XNOR vector; (ii) bitcount of the XNOR vector. This VDP operation is captured in Eq. \ref{eq2}.

\begin{equation}
    z = W \odot I = \sum_{i=1}^{S} W_i \odot I_i
    \label{eq2} 
\end{equation}

Here, $W_i$ and $I_i$, respectively, are the individual bit-elements at index i of the binarized vectors \textit{W} and \textit{I} of size \textit{S} each; $\odot$ denotes the VDP operation (XNOR operation) between binarized vectors \textit{I} and \textit{W} (bit-elements $W_i$ and $I_i$); $\sum$ represents the bitcount operation.

\textbf{Using \{0,1\} instead of \{-1,1\}:} If binary value set \{-1,1\} is used, obtaining the activation values for the next BNN layer after a convolution operation requires $sign(z)$ for each bitcount result $z$. On the other hand, if binary value set \{0,1\} is used, obtaining the activation values for the next BNN layer after a convolution operation requires $compare(z,0.5 \times z_{max})$=$z>0.5\times z_{max}? 1:0$ for each bitcount result $z$, where $z_{max}$ is the size of the binarized vectors \textit{I} and \textit{W}.

\subsection{Processing  of BNNs on Hardware}
\begin{figure}   
    \centering
    \includegraphics[scale=0.59]{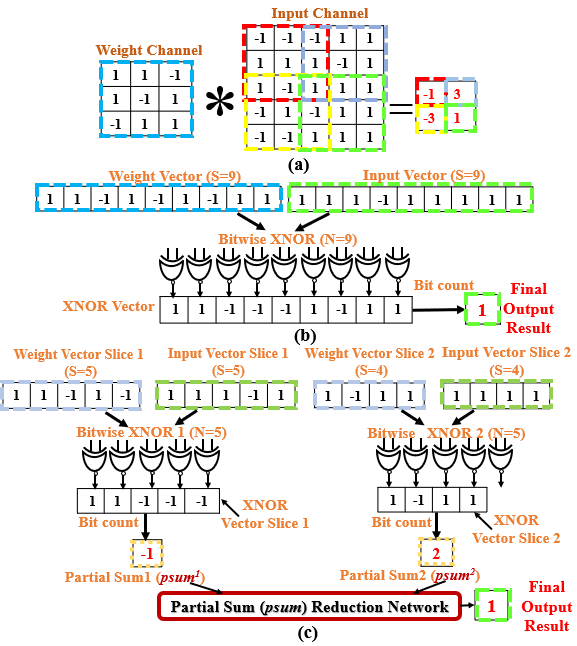}
    \caption{(a) Illustration of a convolution between a weight and input channel in a Binary Neural Network. Bit-wise XNOR and bitcount operations between a flattened weight vector and input vector, (b) when \textit{S}=\textit{N}=9, and (c) when \textit{N}=5, \textit{S}=9; each input and weight vector of \textit{S}=9 is split into two slices (Slice 1 with \textit{S}=5 and Slice 2 with \textit{S}=4). Binary value set \{-1,1\} is used in this example.}
    \label{xnorbitcount}
\end{figure}

Fig. \ref{xnorbitcount}(a) illustrates the convolution between a 3$\times$3 weight channel and a 5$\times$5 input channel. 
During the convolution, based on the stride parameter, the weight channel slides over the input channel and performs inner products with multiple input channel windows (e.g., four input channel windows are shown in Fig. \ref{xnorbitcount}(a) with red, blue, yellow, and green borders), generating one output value per input channel window.   
From Fig. \ref{xnorbitcount}(b), to perform one such inner product (i.e., corresponding to the input channel window highlighted in green in Fig \ref{xnorbitcount}(a)), the input channel window and weight channel are flattened into input and weight vectors of size  \textit{S}=9 each. 
Then, a bitwise XNOR circuit, with a total of \textit{N}=\textit{S}=9 XNOR gates, is employed to generate an XNOR vector. A bitcount circuit then counts the bits in the XNOR vector to evaluate the corresponding inner product output. 
However, the hardware size \textit{N}$\neq$\textit{S} often. 
For example, in Fig. \ref{xnorbitcount}(c), \textit{S}=9 and \textit{N}=5. In this case, both the input and weight vectors (\textit{S}=9 each) are decomposed into two slices each: Slice 1 with \textit{S}=5 and Slice 2 with \textit{S}=4. These slices are then mapped onto two bitwise XNOR circuits with \textit{N}=5 each, as shown in Fig. \ref{xnorbitcount}(c), to consequently produce two XNOR vector slices. Applying bitcount on these XNOR vector slices generates two 
partial sums (\textit{psums}), i.e., \textit{$psum^1$} and \textit{$psum^2$}. \textit{$psum^1$} and \textit{$psum^2$} are then sent to a \textit{psum} reduction network to generate the corresponding inner product output. The addition of the \textit{psums} by the \textit{psum} reduction network incurs additional latency and energy overheads while processing BNNs.    

\begin{figure*}[] 
    \centering
    \includegraphics[scale=0.72]{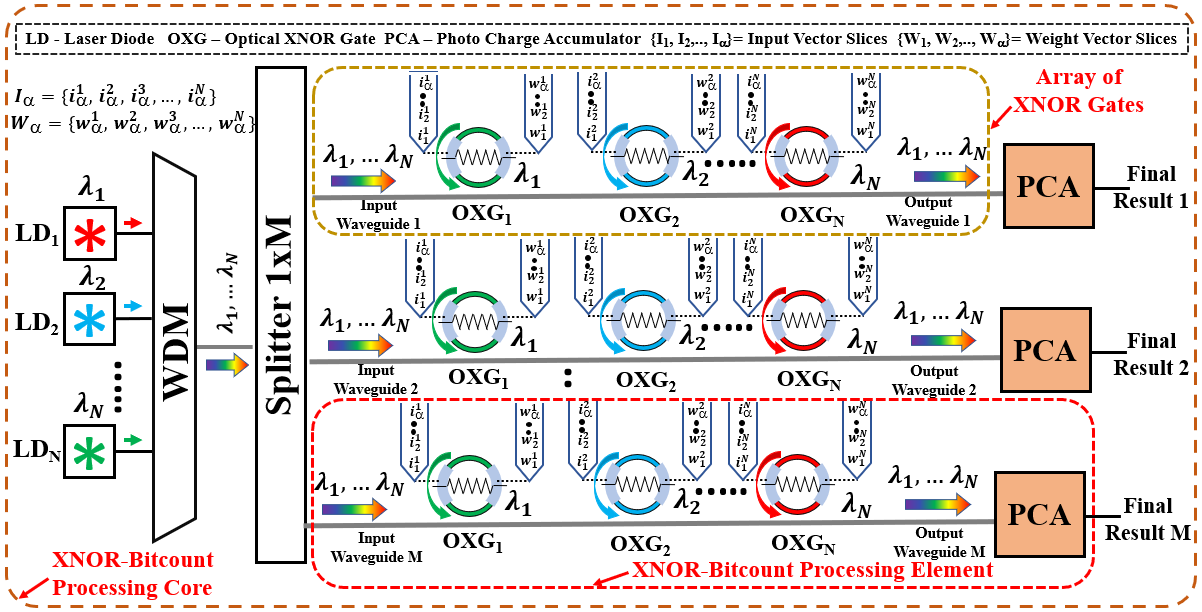}
    \caption{Schematic of an XNOR-Bitcount Processing Core (XPC) of our OXBNN accelerator. Our OXBNN employs binary value set \{0,1\}.}
    \label{OXBNN}
\end{figure*} 
\subsection{Related Work on Optical BNN Accelerators}
To accelerate CNN inferences with low latency and low energy consumption, prior works proposed various accelerators based on photonic integrated circuits (PICs) (e.g., \cite{holylight,crosslight,mzicomplex2021}). These accelerators can be classified as incoherent (e.g., \cite{holylight,crosslight,deapcnn}) or coherent (e.g., \cite{cansu2021,coherent2018}). 
Because of the inherent advantages of incoherent accelerators \cite{crosslight}\cite{cases2022}, the BNN-specific incoherent accelerators \cite{robin} and \cite{lightbulb} were reported. \textit{These optical BNN accelerators from prior works employ binary value set \{0,1\}}. \cite{robin} proposes broadcast and weight styled \cite{karen2020proceeding} XNOR-Bitcount circuits, which use heterogeneous MRRs to mitigate fabrication process variations. In contrast, the microdisk-based accelerator \cite{lightbulb} proposes an all-optical XNOR-Bitcount circuit that uses optical XNOR gates, optical analog-to-digital converters (ADCs), and PCM-based racetrack memory to enable processing at a very high datarate. However, both \cite{robin} and \cite{lightbulb} require at least two MRRs or microdisks to perform a 1-bit XNOR operation (in \cite{lightbulb}, one additional MRR/microdisk is required to modulate the optically applied input operand). Therefore, their XNOR circuits occupy high area and consume high energy.
In addition, the bitcount circuits of these prior works can evaluate only one \textit{psum} at a time by counting the bits of one XNOR vector slice at a time.
Therefore, these circuits have to store the individual \textit{psum}s temporarily in memory. Once sufficient \textit{psum}s are collected, they can be sent to a \textit{psum} reduction network to produce the final result. Thus, the bitcount circuits from prior works incur high memory footprint for storing \textit{psum}s, and high latency and energy for processing \textit{psum}s. 
Our OXBNN accelerator addresses these shortcomings of prior works. 

\section{Our Proposed OXBNN Architecture}
\subsection{Overview}\label{overview}
The main processing unit of our OXBNN architecture is an XNOR-Bitcount Processing Core (XPC), which is illustrated in Fig. \ref{OXBNN}. Our XPC has an array of total \textit{N} single-wavelength laser diodes (LDs), with each LD sourcing optical power of $P_{\lambda_i}^{in}$ amount at a distinct wavelength $\lambda_i$. The total power from all \textit{N} LDs (at wavelengths $\lambda_1$ to $\lambda_N$) multiplex into a single photonic waveguide through wavelength division multiplexing (WDM). The optical power containing all these \textit{N} wavelengths is split into \textit{M} input waveguides, each of which connects to an XNOR-Bitcount Processing Element (XPE) (Fig. \ref{OXBNN}). An XPC contains a total of \textit{M} XPEs.

\subsection{XNOR-Bitcount Processing Element (XPE)}\label{sec3b}
From Fig. \ref{OXBNN}, an XPE in our OXBNN architecture contains two parts: \textit{(i)} an array of a total of \textit{N} Optical XNOR Gates (OXGs) that generates an XNOR vector (or an XNOR vector slice) containing \textit{N} optical bits, and \textit{(ii)} our invented Photo-Charge Accumulator (PCA) that performs bitcount on the generated XNOR vector (or XNOR vector slice). The value \textit{N} here, which is equal to the number of wavelengths and number of OXGs per XPE, is referred to as the size of the XPE. 

\subsubsection{Array of Optical XNOR Gates (OXGs)}
In an XPE, an array of a total of \textit{N} OXGs couples to an input waveguide as shown in Fig. \ref{OXBNN}. Each OXG operates upon a unique wavelength $\lambda_i$ traversing the input waveguide. Each OXG in the array electrically receives two binary operands (i.e., input bit i$^N_1$ and weight bit w$^N_1$) from its corresponding drivers (not shown in the figure). The array of OXGs performs a bit-wise logical XNOR between an \textit{N}-bit input vector slice \textit{$I_1$ = \{$i^1_1,i^2_1,..,i^N_1$\}} and an \textit{N}-bit weight vector slice  \textit{$W_1$ = \{$w^1_1,w^2_1,..,w^N_1  $\}} to produce a resultant \textit{N}-bit XNOR vector slice. Each OXG in the array produces one bit of the resultant XNOR vector slice, and it imprints this bit on its corresponding $\lambda_i$ (by modulating the optical transmission at $\lambda_i$) to be consequently guided to the bitcount circuit (i.e., PCA) via the output waveguide. As a result, the PCA receives the \textit{N} individual optical bits of the \textit{N}-bit XNOR vector slice concurrently on \textit{N} distinct wavelengths. The PCA performs bitcount on these optical bits, as explained later. This entire processing step, from the bit-parallel application of the binary input and weight vector slices at the electrical input terminals of the array of \textit{N} OXGs to the generation of the bitcount result by the PCA, takes very low latency because of the light-speed operation of the XPE. \ul{We refer to this processing step mapped on an XPE as a PASS and the corresponding latency as $\tau$}. Thus, our XPE can produce one bitcount result for one XNOR vector slice in every single PASS with $\tau$ latency. Since $\tau$ can be very low (as low as 20 ps), our XPE can achieve very high processing throughput by completing one PASS every $\tau$ period. For that, multiple input and weight vector slices \{$I_1,I_2,..,I_\alpha$\} and \{$W_1,W_2,..,W_\alpha$\} can be applied to the array of OXGs of an XPE in a serial manner at the predefined data rate (DR) of $\tfrac{1}{\tau}$. The design and operation of an OXG and PCA are explained next.

\begin{figure}[]
    \centering
    \includegraphics[scale=0.33]{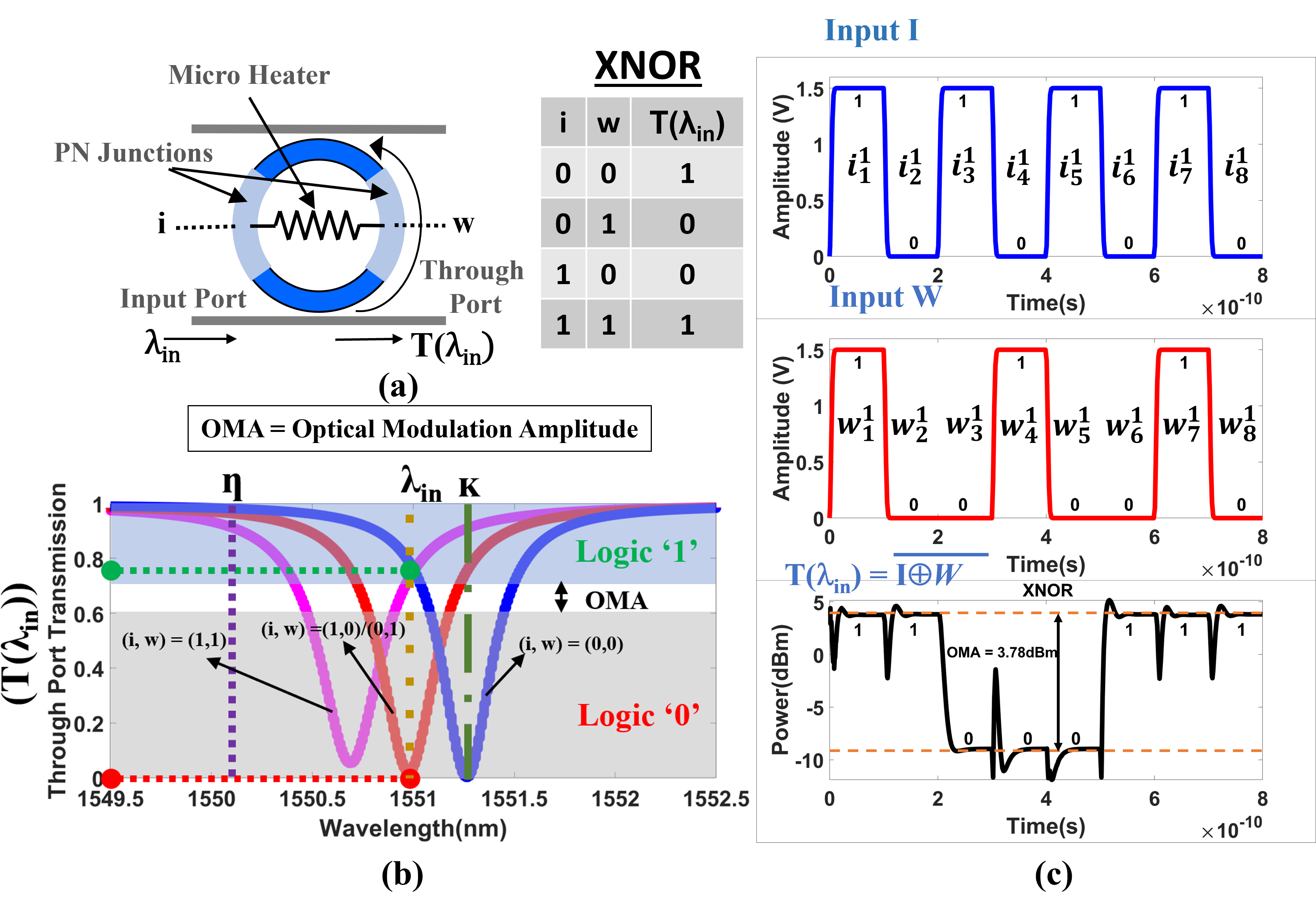}
    \caption{(a) Schematic of our Optical XNOR Gate (OXG). (b) Spectral operation of OXG. (c) Transient analysis of OXG.}
    \label{OXG_Gate}
\end{figure}

\underline{Design of an Optical XNOR Gate (OXG)}: The design of our invented Optical XNOR Gate (OXG) is illustrated in Fig. \ref{OXG_Gate}(a). It is an add-drop microring resonator (MRR), which has two operand terminals (realized as embedded PN-junctions) that can take two operand bits \textit{i} and \textit{w} as inputs for a predefined time-width (usually a little less than the $\tau$ period). Fig. \ref{OXG_Gate}(b) shows the passbands of the MRR for different operand inputs and temperature conditions. The MRR's temperature can be increased using the integrated microheater (Fig. \ref{OXG_Gate}(a)), to consequently tune its operand-independent resonance from its fabrication-defined initial position \textit{$\eta$} to its programmed position \textit{$\kappa$} (blue passband; Fig. \ref{OXG_Gate}(b)), relative to the input optical wavelength position $\lambda_{in}$. For each bit combination at the operand terminals ((\textit{i},\textit{w}) = (0,1), (1,0), or (1,1)), the MRR's resonance passband electro-refractively moves to an operand-driven position (red and magenta passbands in Fig. \ref{OXG_Gate}(b)). Based on the MRR resonance passband's programmed position \textit{$\kappa$} relative to $\lambda_{in}$, the through-port transmission (T($\lambda_{in}$)) of the MRR provides bit-wise logical XNOR operation between the input bits \textit{i} and \textit{w}.

To validate the operation of our OXG, we performed the transient analysis, as shown in Fig. \ref{OXG_Gate}(c). For that, we modelled and simulated our OXG using the foundry-validated tools from Ansys/Lumerical's DEVICE, CHARGE, and INTERCONNECT suites \cite{lumerical_2021}. Fig. \ref{OXG_Gate}(c) shows two input bit-streams $I$ = \{$i^1_1,i^1_2,..,i^1_8$\} and $W$ = \{$w^1_1,w^1_2,..,w^1_8$\} applied to the two PN junctions of our OXG at a DR = 10 GS/s. By looking at the output optical trace  T($\lambda_{in}$) in  Fig. \ref{OXG_Gate} (c), we can say T($\lambda_{in}$) = \{$i^1_1 \odot w^1_1$,..,$i^1_8 \odot w^1_8$\}, which validates the functionality of our OXG as a logical XNOR gate.\ul{
From our validation, our OXG has a full passband width at half maximum (FWHM) of 0.35 nm and it can operate at DR of up to 50 GS/s. Our XNOR gate consumes energy of 0.032nJ with an area footprint of 0.011mm$^2$}.

\subsubsection{Photo-Charge Accumulator (PCA)}\label{secpca}
\begin{figure}[] 
    \centering
    \includegraphics[scale=0.58]{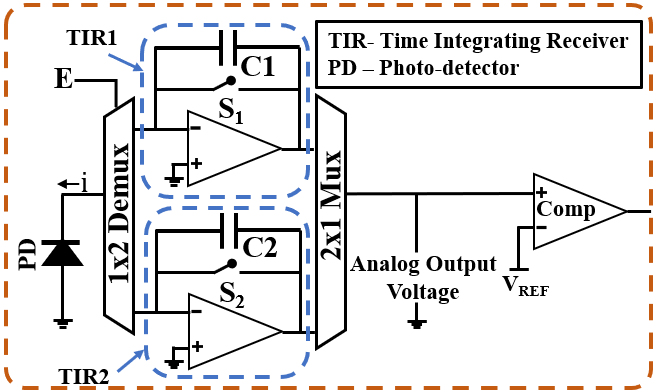}
    \caption{Photo-Charge Accumulator (PCA) Circuit. $V_{REF}$ is the threshold required in the $compare()$ function discussed in Section \ref{section2a}. Typically, $V_{REF}$ = 2.5V because we consider the dynamic range of TIR to be 5V.}
    \label{PCA}
\end{figure}

From Section \ref{overview}, the XNOR vector bits generated by an array of OXGs are guided to a PCA circuit, where a bitcount is performed on the XNOR vector bits to generate an output result. Our PCA circuit employs a photodetector and two time integrating receiver (TIR) circuits \cite{alexandermit2022} (one of the TIR1 and TIR2 circuits remains redundant, enabled by the demux and mux; Fig \ref{PCA}). The photodetector generates a current pulse for each optical logic `1' incident upon it. The amplitude of a current pulse generated for an optical logic '0' remains under the noise limit; therefore, a logic '0' remains statistically undetected. The current pulse generated by an optical logic '1' accumulates a certain statistically significant amount of charge on the capacitor of the active TIR circuit (e.g., the circuit with C1 capacitor); as a result, the TIR circuit outputs a detectable analog voltage level \cite{alexandermit2022}. Hence, when more optical '1's are incident upon the photodetector, the total accumulated charge on the active capacitor (e.g., C1), and thus, the accrued output analog voltage level, grows proportionally to the total number of optical `1's that are incident \cite{alexandermit2022}. This is because a current source (a sequence of current pulses) can charge a capacitor linearly following this equation: $\delta V$=$\tfrac{i \delta t}{C}$, where $i$ is an incident current pulse, $\delta t$ is the time-width of the current pulse, $C$ is the capacitance, and $\delta V$ is the accrued voltage. The final analog voltage accrued at the TIR output, thus, represents the bitcount result (accumulation result) of the incident optical '1's. However, the number of '1's that can be accumulated in such a manner might be limited, as the output of the TIR circuit (Fig. \ref{PCA}) might saturate. Once the output of a TIR circuit saturates, the ongoing accumulation phase ends and the bitcount result (i.e., the final TIR output voltage) is passed through a comparator to generate the activation value for the next BNN layer (as explained in Section \ref{section2a}).
After one accumulation phase, a  discharge of the active capacitor (e.g., C1) is needed to prepare the circuit for the next accumulation phase. While capacitor C1 is discharging, the redundant TIR2 circuit with capacitor C2 mitigates the discharge latency by allowing a continuation of a concurrent bitcount.  
          
\section{Scalability Analysis and Mapping}\label{sec4}

\subsection{Scalability of XNOR-Bitcount Processing Cores (XPCs)}\label{secscalability}
To determine the achievable size \textit{N} for our XPC, we adopt scalability analysis equations (Eq. \ref{eq3}, Eq. \ref{eq4}, and Eq. \ref{eq5}) from \cite{lukas} and \cite{cases2022}. Table \ref{abbrevations} reports the definitions of the parameters and their values used in these equations. 
  We considered Free Spectral Range (FSR=50nm) \cite{lukas}, FWHM=0.35nm (refer Section \ref{sec3b}), and inter-wavelength gap of 0.7nm. For these spectral conditions, we observed minimal crosstalk power penalty for the OXGs operating at DR=50GS/s ($<$1 dB penalty \cite{karenbergman,praneethsos,proteus}, which is accounted for as part of parameter $IL_{penalty}$ in the equations (Table \ref{abbrevations})). Since the XPC of our OXBNN accelerator processes binarized vectors, it requires the bit precision of \textit{B}=1-bit in the equations. We consider \textit{M=N} and first solve Eq. \ref{eq3} and Eq. \ref{eq4} for a set of DRs=\{3, 5, 10, 20, 30, 40, 50\} GS/s, to find a corresponding set of $P_{PD-opt}$. Then, we solve Eq. \ref{eq4} for \textit{N} with the obtained set of $P_{PD-opt}$ values across the set of \textit{DR}s. Table \ref{scalability} reports the achievable \textit{N} for our XPC across various \textit{DR}s. As evident, the supported \textit{N} value decreases from \textit{N=66} at 3 GS/s to \textit{N=19} at 50 Gs/s. This achievable \textit{N} value defines the feasible number of OXGs per XPE; thus, this \textit{N} also defines the maximum size of the XNOR vector slice that can be generated in our XPC. Because we consider FSR of 50nm and inter-wavelength gap of 0.7nm, we verify that the maximum \textit{N}=66 can be supported within the FSR (i.e., \textit{N}=66$<$(FSR/0.7nm)).

\begin{equation}
       B = \frac{1}{6.02}\Bigg[20log_{10}(\frac{R_{s}\times P_{PD-opt}}{\beta\sqrt{\frac{DR}{\sqrt{2}}}}-1.76\Bigg]
       \label{eq3}
\end{equation}

\begin{equation}
    \beta = \sqrt{2q(R_{s}P_{PD-opt}+I_d)+\frac{4kT}{R_L}+R^2_{s}P_{PD-opt}^2RIN}
    \label{eq4}
\end{equation}

\begin{equation}
   \label{eq5}
\begin{split}
 P_{Laser} = \frac{10^{\frac{\eta_{WG}(dB)[N(d_{OXG})+d_{element}]}{10}}M}{\eta_{SMF}\eta_{EC}\eta_{WPE}IL_{i/p-OXG}}
    \times\frac{P_{PD-opt}}{IL_{penalty}}
      \\\times\frac{1}{(OBL_{OXG})^{N-1}(EL_{splitter})^{log_{2}M}} 
    \\ 
\end{split}
\end{equation}

 \underline{Analysis of PCA's Accumulation Capacity}: 
 We modeled the photodetector (PD) of our PCA circuit using the INTERCONNECT tool from Ansys/Lumerical \cite{lumerical_2021} for PD responsivity = 1.2 A/W across different $P_{PD-opt}$ values corresponding to the \textit{N} values in Table \ref{scalability}. We extracted the current pulse values generated by the photodetector for the incident optical `1's and `0's corresponding to each $P_{PD-opt}$. We then imported these values in our MultiSim \cite{multisim} based model of the PCA with C1=C2=10pF \cite{alexandermit2022}, and the TIR gain=50. For these parameters, we simulated the analog output voltage at the PCA's TIR for different bitcount results (i.e., different values of the total number of accumulated `1's). From this analysis, we observed that the maximum number of '1's that can be accumulated by our PCA is limited by the available operating dynamic range of the TIR of our PCA. We considered the TIR's operating dynamic range to be 5V (0V to 5V) and evaluated our PCA's accumulation capacity $\gamma$, which we define as the maximum number of `1's  that can be accumulated by the PCA within the TIR's operating dynamic range. Our evaluated $\gamma$ values, for each pair of \textit{N} and corresponding $P_{PD-opt}$, are reported in Table \ref{scalability}. Since our PCA can accumulate a total of $\gamma$ bits and since each XNOR vector slice in our XPC has a total of \textit{N} bits, our PCA can accumulate a total of $\alpha$ XNOR vector slices, where $\alpha$ = $\frac{\gamma}{N}$. Table \ref{scalability} also reports the values of $\alpha$. As evident, the $\gamma$ and $\alpha$ values for our PCA can be very large, which provides several substantial benefits as discussed in Section \ref{sec4c}. 

\begin{table}[]
\centering
\caption{Definition and values of various parameters used in Eq. \ref{eq3}, Eq. \ref{eq4}, and Eq. \ref{eq5} (from \cite{lukas}) for the scalability analysis. Definitions of PCA parameters $\gamma$ and $\alpha$.}
\label{abbrevations}
\begin{tabular}{|c|c|c|}
\hline
{ \textbf{Parameter}}             & { \textbf{Definition}}                                                                                                         & { \textbf{Value}}   \\ \hline
{$P_{Laser}$=$P_{\lambda_i}^{in}$}                & { Laser Power Intensity}                                                                                               & { 5 dBm}  \\ \hline
 { $R_{s}$}                     & { PD Responsivity}                                                                                                     & { 1.2 A/W} \\ \hline
{ $R_L$}                    & { Load Resistance}                                                                                                     & { 50 $\Omega$}      \\ \hline
{ $I_d$}                    & { Dark Current}                                                                                                        & { 35 nA}   \\ \hline
 { T}                     & { Absolute Temperature}                                                                                                & { 300 K}   
\\ \hline
{ RIN}                   & { Relative Intensity Noise}                                                                                            & { -140 dB/Hz}   \\ \hline
 { $\eta_{WPE}$}              & { Wall Plug Efficiency}                                                                                                & { 0.1}   \\ \hline  
{ $IL_{SMF}$}           & { Single Mode Fiber Insertion Loss}                                                                                    & { 0 dB}       \\ \hline
{ $IL_{EC}$}            &\begin{tabular}[c]{@{}c@{}}Fiber to Chip Coupling \\ Insertion Loss\end{tabular}                                                                                & { 1.6 dB}     \\ \hline
{ $IL_{WG}$}         & \begin{tabular}[c]{@{}c@{}} Silicon Waveguide  \\ Insertion Loss\end{tabular}                                                                                   & { 0.3 dB/mm}     \\ \hline
 { $EL_{Splitter}$}      & { Splitter Insertion Loss}                                                                                             & { 0.01 dB}    \\ \hline
 { $IL_{OXG}$}           & \begin{tabular}[c]{@{}c@{}}Optical XNOR Gate (OXG)  \\ Insertion Loss\end{tabular}                                                                           & { 4 dB}       \\ \hline
 { $OBL_{OXG}$}          & { Out of Band Loss OXG}                                                                                                & { 0.01 dB}                                                                                 \\ \hline
 { $IL_{penalty}$} & { Network Penalty}                                                                                                     & { 4.8 dB}     \\ \hline
 { $d_{OXG}$}                & { Gap between two adjacent OXGs}                                                                                       & { 20 $\mu$m} \\ \hline 
 
 { $P_{PD-opt}$}                & { Output Photodetector Sensitivity}                                                                                       & {Table \ref{scalability}} \\ \hline
 
 { $\alpha$}                & \begin{tabular}[c]{@{}c@{}}PCA's Accumulation Capacity  \\ (\# of XNOR Vectors Slices) \end{tabular}                                                                                         & {Table \ref{scalability}} \\ \hline 
 { $\gamma$}                & \begin{tabular}[c]{@{}c@{}}PCA's Accumulation Capacity  \\ (\# of accumulated `1's) \end{tabular}                                                                                         & {Table \ref{scalability}} \\ \hline 
 
\end{tabular}
\end{table}
\begin{table}[]
\centering
\caption{XPC Size \textit{N}, PCA bitcount capacity values ($\gamma$ and $\alpha$), for different data rates (DRs).}
\label{scalability}
\begin{tabular}{|c|c|c|c|c|}
\hline
\textbf{Datarate (DR) (GS/s)} & \textbf{$P_{PD-opt}(dBm)$} & \textbf{N} & \textbf{$\gamma$} &  \textbf{$\alpha$} \\ \hline
3                 & -24.69                        & 66  &  39682
     & 601            \\ \hline
5                 & -23.49                        & 53  &   29761
   & 561            \\ \hline
10                & -21.9                         & 39  &   19841
    & 508            \\ \hline
20                & -20.5                         & 29  &   14880
   & 513            \\ \hline
30                & -19.5                         & 24  &    10822
   & 450            \\ \hline
40                & -18.9                         & 21  &  9920
    & 472            \\ \hline
50                & -18.5                         & 19   &  8503
    & 447            \\ \hline
\end{tabular}
\end{table}

\subsection{Mapping Convolutions on an XPC}\label{secxvdpcop}
As described in Section \ref{section2a}, for processing a BNN convolution on hardware, both the weight and input channels are flattened into binarized vectors. For mapping of a binary convolution on an XPC (or XPE), these binarized input and weight vectors are represented as matrices. For instance, the input matrix  \textit{$\mathbb{I}$(H, S)} has \textit{H} rows corresponding to \textit{H} binarized input vectors of size \textit{S} each. 
Similarly, the weight matrix \textit{$\mathbb{W}$(H, S)} can also be defined. These matrices \textit{$\mathbb{W}$(H, S)} and \textit{$\mathbb{I}$(H, S)} are mapped onto an XPC containing a total of \textit{M} XPEs of size \textit{N} each. Depending on the relation between \textit{S} and  \textit{N}, two cases drive the selection of the appropriate mapping. These cases and their corresponding mappings are illustrated in Fig. \ref{OXBNNOperation}, \underline{for \textit{M}=2, \textit{H}=2, \textit{N}=9, and two distinct values of \textit{S}}. These cases are explained below: 

\begin{figure}[]
    \centering
        \includegraphics[scale=0.45]{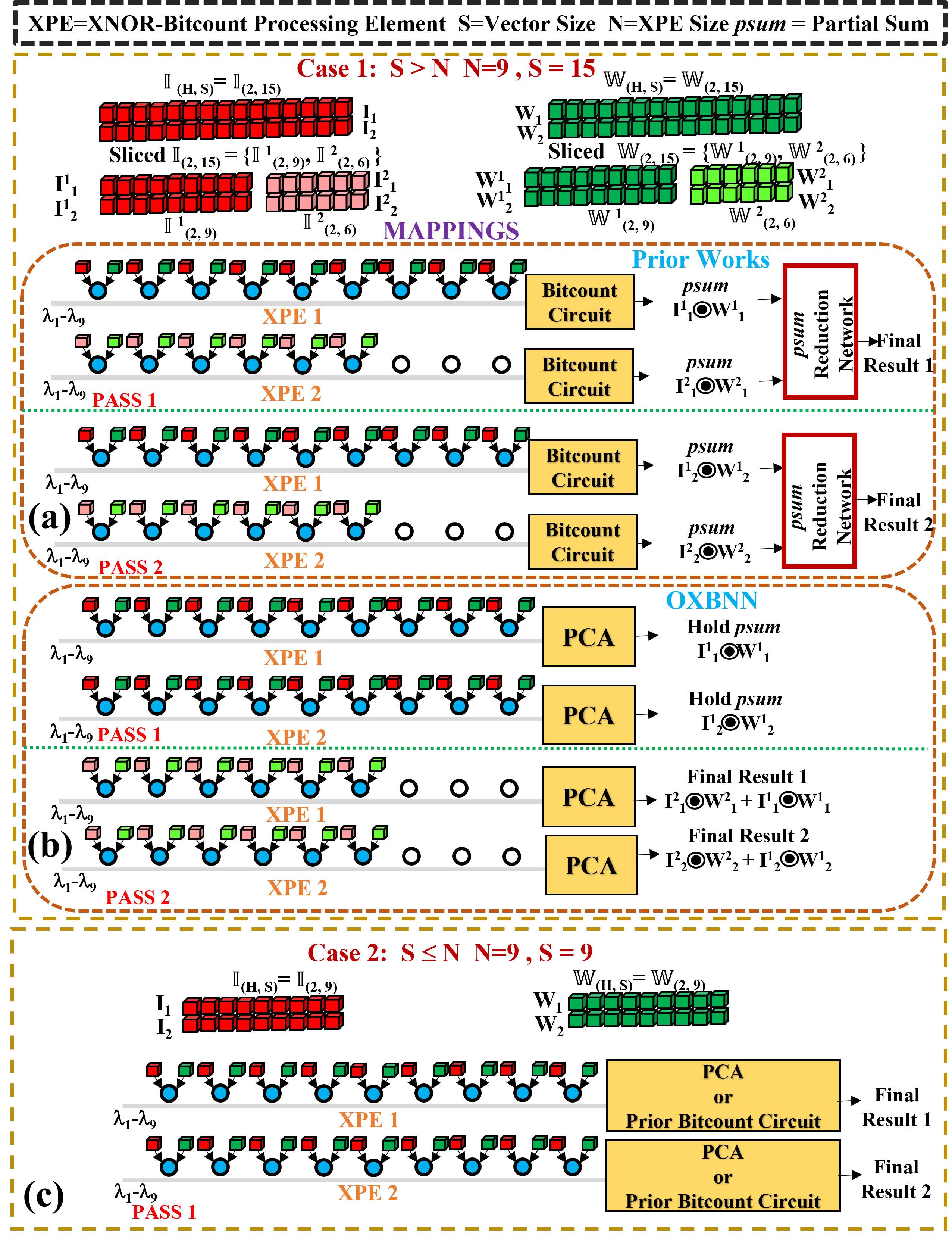}
    \caption{
    Example mappings and related operation of our XPC for various cases of the \textit{S} and \textit{N} values. A comparison of our PCA with the bitcount circuit from prior works is also illustrated.}
    \label{OXBNNOperation}
\end{figure}

\underline{\textbf{Case 1, \textit{S}=15, \textit{S}$>$\textit{N}, Fig. \ref{OXBNNOperation}(a) and \ref{OXBNNOperation}(b)}}: Matrices $\mathbb{I}$ and $\mathbb{W}$ consist of two vectors each, \{$I_1$, $I_2$\} and  \{$W_1$, $W_2$\}, respectively. To make the size \textit{S=15} of these vectors \{$I_1$, $I_2$\} and \{$W_1$, $W_2$\} amenable to the XPE size \textit{N=9}, each of these vectors is split into two slices to yield a set of input vector slices \{$I_1^1$, $I_1^2$, $I_2^1$, $I_2^2$\} and a set of weight vector slices \{$W_1^1$, $W_1^2$, $W_2^1$, $W_2^2$\}. Since \textit{M=2} is less than the total number of vector slices (i.e., $H\times ceil(S/N)$ = 4), multiple passes are required to complete the processing of these vector slices. Mappings of these vector slices differ between our PCA and the bitcount circuit from prior works \cite{robin} and \cite{lightbulb}, as discussed next.   

\textbf{Mapping for the bitcount circuit from \cite{robin} and \cite{lightbulb} (Fig. \ref{OXBNNOperation}(a))}: Since M=2, there are two XPEs, namely XPE 1 and XPE 2. During PASS 1 of these XPEs (the definition of a PASS is given in Section \ref{sec3b}), we map \{$I^1_1$, $W^1_1$\} onto XPE 1, and \{$I^2_1$, $W^2_1$\} onto XPE 2. XPE 1 generates the corresponding XNOR vector, which is accumulated using the bitcount circuit to produce \textit{psum} $I^1_1 \odot W^1_1$. Similarly, XPE 2 generates \textit{psum} $I^2_1 \odot W^2_1$. The generated \textit{psums} are reduced (further accumulated) at the \textit{psum} reduction network, to produce Final Result 1. Similarly, during PASS 2, vector slices 
\{$I^1_2$, $I^2_2$, $W^1_2$, $W^2_2$ \} are mapped to generate corresponding \textit{psum}s, which are then sent to the \textit{psum} reduction network to produce Final Result 2. Thus, for the bitcount circuits from prior works, there is a need for employing a \textit{psum} reduction network, which leads to a high latency overhead. 

\textbf{Mapping for our OXBNN with PCAs, Fig. \ref{OXBNNOperation}(b)}: Our OXBNN maps all the slices of a particular vector to the same XPE. During PASS 1, OXBNN maps \{$I^1_1$, $W^1_1$\} to XPE 1, and \{$I^1_2$, $W^1_2$\} to XPE 2. XPE 1 charges its PCA's capacitor to generate an analog voltage level that represents \textit{psum} $I^1_1 \odot W^1_1$, whereas XPE 2 charges its PCA's capacitor to generate an analog voltage level that represents \textit{psum} $I^1_2 \odot W^1_2$. Because a PCA can accumulate a total of $\alpha$ vector slices (Section \ref{secpca}), the PCAs of XPE 1 and XPE 2 can be made to hold the charge and analog voltage accrued during PASS 1. Then, during PASS 2, XPE 1 and XPE 2 can further grow these held analog voltage levels by the amounts proportional to $I^2_1 \odot W^2_1$  and $I^2_2 \odot W^2_2$, respectively. Thus, at the end of PASS 2, the total accrued analog voltage on the PCA of XPE 1 (XPE 2) would be proportional to $I^1_1 \odot W^1_1$ + $I^2_1 \odot W^2_1$ ($I^1_2 \odot W^1_2$ + $I^2_2 \odot W^2_2$). Thus, the PCAs of our OXBNN can accumulate multiple \textit{psums} (a total of $\alpha$ \textit{psums}) inherently. This eliminates the need to employ \textit{psum} reduction networks to consequently yield substantial benefits, as further explained in Section \ref{sec4c}. 

\underline{\textbf{Case 2, \textit{S}=9, \textit{S}$\leq$\textit{N}, Fig. \ref{OXBNNOperation} (c)}}:
The size \textit{S=9} of the vectors \{$I_1$, $I_2$\} and \{$W_1$, $W_2$\} matches with the XPE size \textit{N=9}. Thus, in a single pass (PASS 1), our OXBNN maps \{$I_1$, $W_1$\} to XPE 1, and \{$I_2$, $W_2$\} to XPE 2. XPE 1 and XPE 2 produce Final Result 1 and Final Result 2 corresponding to $I_1 \odot W_1$ and $I_1 \odot W_1$, respectively. In this case, the mapping is identical for our PCA and the bitcount circuits from prior work.

\subsection{Latency and Energy Benefits of PCA} \label{sec4c}
Our PCA provides manifold benefits in terms of both the latency and energy consumption. The latency benefits accrue because our PCA eliminates the need of employing \textit{psum} reduction networks to temporarily store and accumulate \textit{psums}. From Section \ref{sec4} and Table \ref{scalability}, our PCA can achieve $\gamma$=8503 and $\alpha$=447 at $DR$=50 GS/s, which means that our PCA, before it saturates, can accumulate a total of $\gamma$=8503 '1's across a total of $\alpha$=447 XNOR vector slices.
As a result, if we operate the OXGs of our OXBNN at $DR$=50 GS/s, our PCA can inherently accumulate (perform bitcount on) any XNOR vector whose size \textit{S} is less than $\gamma$=8503. Since the maximum XNOR vector size is observed to be $S$=4608 across all major modern CNNs (e.g., ResNet18, ResNet50, DenseNet121, VGG16, VGG19, GoogleNet, Inception\_V3, EffiecientNet\_B7, NASNetMobile, MobileNet\_V2, and ShuffleNet) \cite{chollet2015keras},
our PCA eliminates the need to employ dedicated \textit{psum} reduction networks in our OXBNN accelerator.


\section{Evaluation}
\subsection{System-Level Implementation of OXBNN.}
\begin{figure}
    \centering
    \includegraphics[scale=0.39]{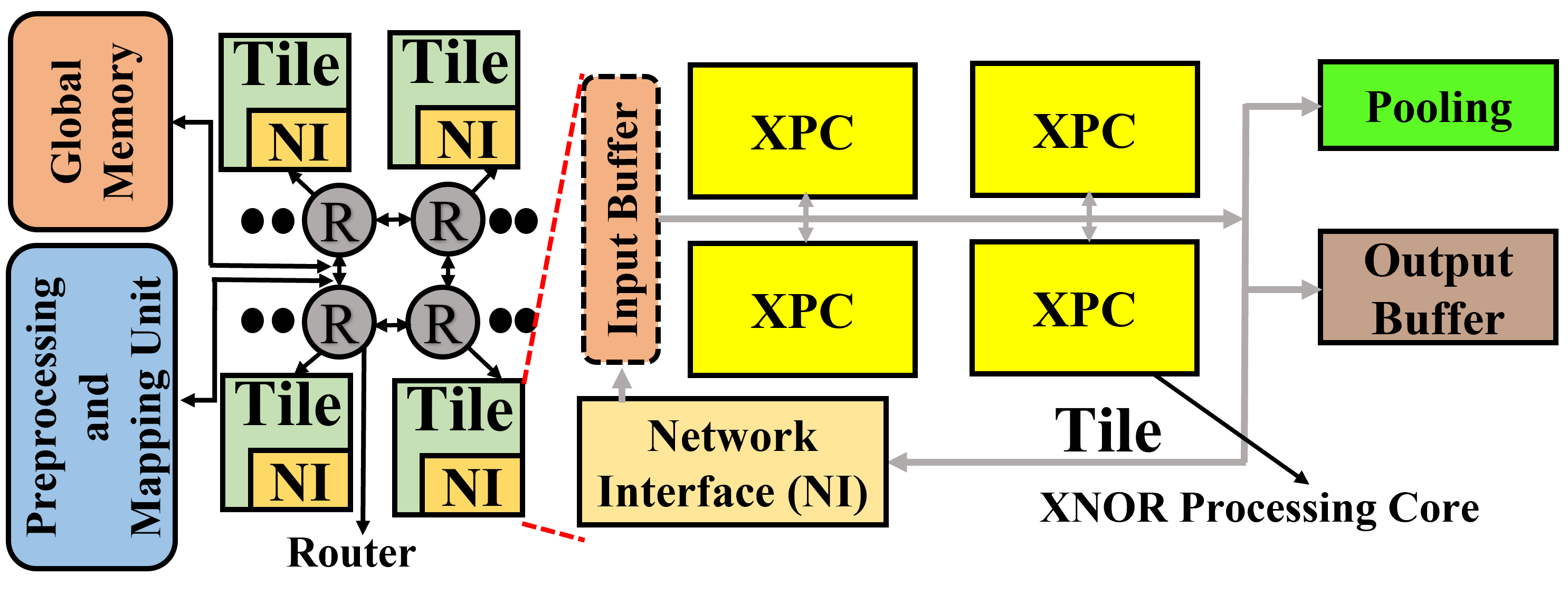}
    \caption{System-level overview of our OXBNN accelerator.}
    \label{syslevelimp}
\end{figure}

Fig. \ref{syslevelimp} illustrates the system-level implementation of our OXBNN accelerator. It consists of global memory that stores BNN parameters and a pre-processing and mapping unit. It has a mesh network of tiles. Each tile contains 4 XPCs interconnected (via H-tree) with an output buffer as well as pooling units.

\subsection{Simulation Setup}
For evaluation, we model our OXBNN accelerator from Fig. \ref{syslevelimp} using our custom-developed, transaction-level, event-driven python-based simulator (\url{https://github.com/uky-UCAT/B\_ONN\_SIM}). We simulated the inference of four BNNs (batch size=1): VGG-small\cite{lqnets}, ResNet18\cite{resnet}, MobileNet\_V2 \cite{mobilenetv2}, and ShuffleNet\_V2 \cite{shufflenet}. We binarized all the weights and inputs using the LQ-Nets technique\cite{lqnets}. We evaluate frames-per-second (FPS) and FPS/W (energy efficiency).  

We compared our OXBNN with ROBIN \cite{robin} and LIGHTBULB \cite{lightbulb}. ROBIN and LIGHTBULB operate at different DRs; therefore, we consider two variants of our OXBNN: (1) OXBNN\_5 with DR=5GS/s (matching with ROBIN) and \textit{N=53} (Table \ref{scalability}), (2) OXBNN\_50 with DR=50GS/s (matching with LIGHTBULB) and \textit{N=19} (Table \ref{scalability}). We consider two variants of ROBIN: ROBIN Energy-Optimized (ROBIN\_EO) and ROBIN Performance-Optimized (ROBIN\_PO)\cite{robin}. For fair comparison, we perform area proportionate analysis, wherein we altered the XPE count for each photonic BNN accelerator across all of the accelerator's XPCs to match with the area of OXBNN\_5 having 100 XPEs. Accordingly, the scaled XPE counts of OXBNN\_50 (\textit{N=19}), ROBIN\_PO (\textit{N=50}), ROBIN\_EO (\textit{N=10}), and LIGHTBULB (\textit{N=16}) are 1123, 183, 916, and 1139, respectively. Table \ref{table3} gives the parameters used for our evaluation.

\begin{table}[]
\caption{Accelerator Peripherals and XPE Parameters {\cite{cases2022}}}
\label{table3}
\begin{tabular}{|c|ccc|}
\hline
                           & \multicolumn{1}{c|}{\textbf{Power(mW)}} & \multicolumn{1}{c|}{\textbf{Latency}} & \textbf{Area($mm^2$)} \\ \hline
\textbf{Reduction Network} & \multicolumn{1}{c|}{0.050}              & \multicolumn{1}{c|}{3.125ns}          & 3.00E-5            \\ \hline
\textbf{Activation Unit}   & \multicolumn{1}{c|}{0.52}               & \multicolumn{1}{c|}{0.78ns}           & 6.00E-5            \\ \hline
\textbf{IO Interface}      & \multicolumn{1}{c|}{140.18}             & \multicolumn{1}{c|}{0.78ns}           & 2.44E-2            \\ \hline
\textbf{Pooling Unit}      & \multicolumn{1}{c|}{0.4}                & \multicolumn{1}{c|}{3.125ns}          & 2.40E-4            \\ \hline
\textbf{eDRAM}             & \multicolumn{1}{c|}{41.1}               & \multicolumn{1}{c|}{1.56ns}           & 1.66E-1             \\ \hline
\textbf{Bus}               & \multicolumn{1}{c|}{7}                  & \multicolumn{1}{c|}{5 cycles}         & 9.00E-3               \\ \hline
\textbf{Router}            & \multicolumn{1}{c|}{42}                 & \multicolumn{1}{c|}{2 cycles}         & 1.50E-2              \\ \hline
\textbf{EO Tuning}         & \multicolumn{1}{c|}{80 $\mu$W/FSR}          & \multicolumn{1}{c|}{20ns}             & -                  \\ \hline
\textbf{TO Tuning}         & \multicolumn{1}{c|}{275 mW/FSR}         & \multicolumn{1}{c|}{4$\mu$s}             & -                  \\ \hline

\end{tabular}

\end{table}

\subsection{Evaluation Results}

\begin{figure}[h!]
  \centering
  \includegraphics[width=\linewidth]{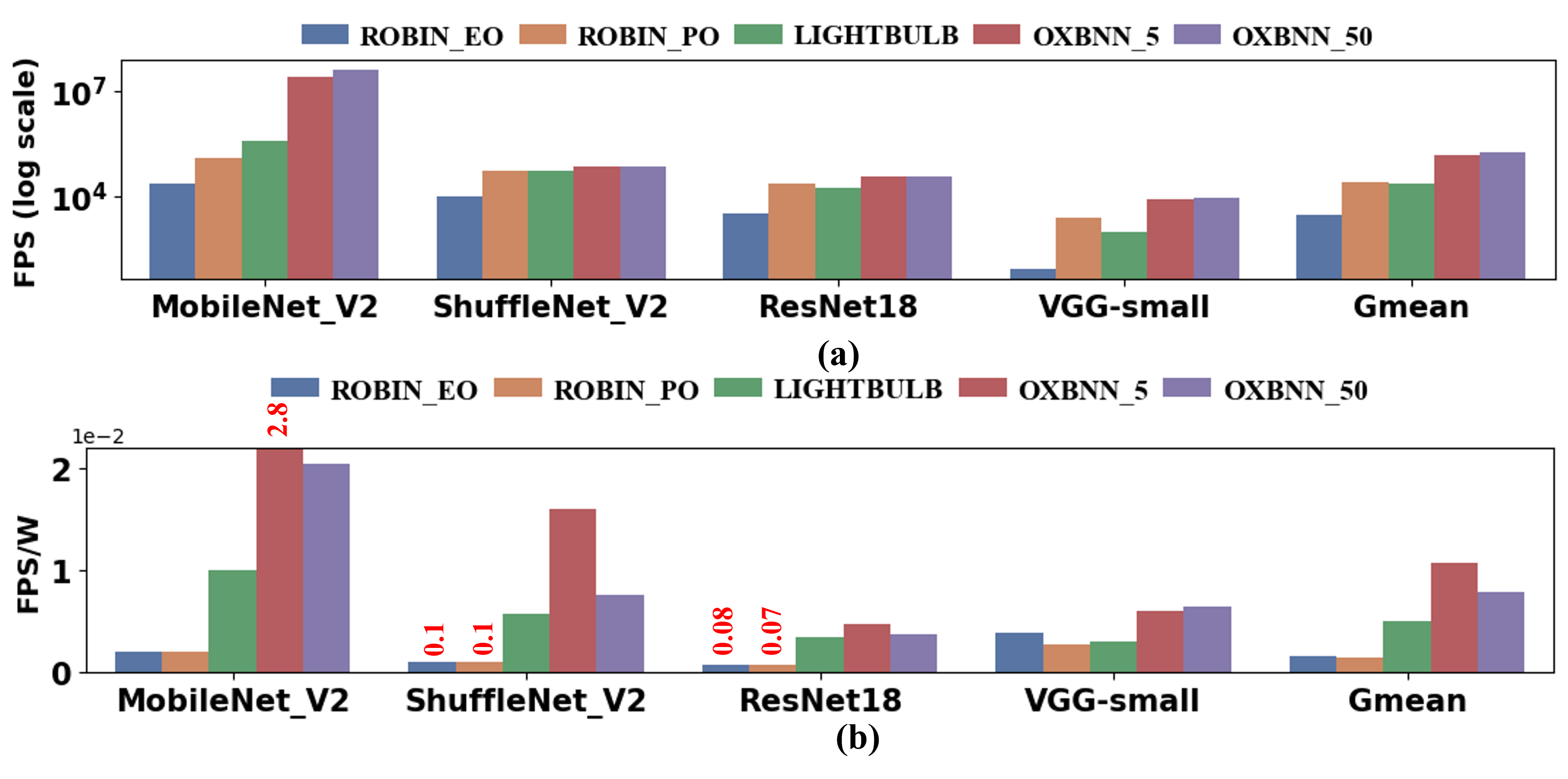}
  \caption{(a) FPS (log scale) (b) FPS/W for OXBNN versus ROBIN and LIGHTBULB accelerators.} 
  \label{FPS}
\end{figure}

Fig. \ref{FPS}(a) compares FPS values (log scale). OXBINN\_50 achieves 62$\times$, 8$\times$, and 7$\times$ better FPS than ROBIN\_EO, ROBIN\_PO, and LIGHTBULB, respectively, on gmean across the BNNs. Similarly, OXBNN\_5 also outperforms 
ROBIN\_EO, ROBIN\_PO, and LIGHTBULB by 54$\times$, 7$\times$, and 16$\times$, respectively, on gmean across the BNNs. OXBNN\_5 and OXBNN\_50 mainly benefit from the elimination of psum reduction networks. They also benefit from their larger \textit{N}, compared to the other accelerators. Larger \textit{N} renders them with higher parallelism and lower $\alpha$ to consequently increase (decrease) their processing throughput (latency).

Fig. \ref{FPS}(b) gives the energy efficiency (FPS/W) achieved by each accelerator across various BNNs. Our OXBINN\_5 gains 6.8$\times$, 7.6$\times$, and 2.14$\times$ better FPS/W than ROBIN\_EO, ROBIN\_PO, and LIGHTBULB, respectively, on gmean across the BNNs. Similarly, our accelerator OXBNN\_50 also outperforms 
ROBIN\_EO, ROBIN\_PO, and LIGHTBULB by 4.9$\times$, 5.5$\times$, and 1.5$\times$, respectively, on gmean across the BNNs. The energy benefits of OXBNN\_5 and OXBNN\_50 are due to the novel OXGs. Due to their single-MRR design, these OXGs consume less energy and static power, compared to the OXGs (containing at least two MRRs or microdisks per OXG) from ROBIN and LIGHTBULB. Moreover, the elimination of the dedicated \textit{psum} reduction network (Section \ref{sec4c}) also eliminates related high energy consumption. Thus, these benefits collectively render better FPS/W for OXBNN\_5 and OXBNN\_50.

\section{Conclusions}
In this paper, we present a single-MRR-based optical XNOR gate (OXG) and a novel bitcount circuit Photo-Charge Accumulator (PCA). We employ OXGs and PCAs to forge a novel accelerator, called 
OXBNN, to process the inferences of BNNs. We performed a comprehensive analysis to show the throughput and energy efficiency advantages of OXBNN. Our evaluation results show that OXBNN provides improvements of up to 62$\times$ and  7.6$\times$ in throughput (FPS) and energy efficiency (FPS/W), respectively, on geometric mean over two state-of-the-art photonic BNN accelerators from prior works.

\section*{Acknowledgments}
We thank the anonymous reviewers whose valuable feedback helped us improve this paper. We would also like to acknowledge the National Science Foundation (NSF) as this research was supported by NSF under grant CNS-2139167.

\bibliographystyle{IEEEtran}
\bibliography{references}

\end{document}